\documentclass{ws-procs9x6}

\def\GeV{\rm GeV}
\def\msbar{\overline{\rm MS}}

\begin{document}

\title{The Structure Functions and Low $x$ ~~~~~~~Working Group Summary}

\author{\vspace{-0.8cm}V. Chekelian$^*$ (Shekelyan), 
C. Gwenlan$^{**}$ and R.S. Thorne$^{***}$}

\address{$^*$Max Planck Institute for Physics, Munich, Germany \\
$^{**}$University of Oxford, UK\\ 
$^{***}$ University College London, UK}

\maketitle

\abstracts{\vspace{-1.1cm}The structure functions and low 
$x$ working group summary of DIS06.}

\vspace{-1.9cm}

\section{Recent results from HERA}

\vspace{-0.1cm}

We begin by discussing new results from HERA. 
In autumn 2003 the accelerator started the second
phase of its $ep$ collider programme (HERA-II). 
The $e^+p$ and $e^-p$ data collected 
by the H1 and ZEUS experiments since then, were taken with longitudinally
polarised positron and electron beams for the first time. 
Recent results from HERA-II, related mostly to the exploration of 
this new feature of the collider, are discussed below. 
The other HERA result presented was an update of 
the high-$x$ measurement of the NC cross sections by the 
ZEUS collaboration\cite{highx}. In a discussion session devoted to
the longitudinal structure function, $F_L$, it was highlighted that as a test 
of different theoretical models a measurement of $F_L(x,Q^2)$ over as 
wide a range of 
$x$ and $Q^2$ as possible would be very useful\cite{Ringberg},
and both H1 and ZEUS expressed a strong interest in
running HERA at a low proton energy for about three months, 
which would allow a direct $F_L$ measurement\cite{fl}.
The experimental and theoretical status of
the charm and beauty contributions to $F_2$ are summarised in the 
Heavy Flavour working group summary.  


Measurements of charged current (CC) deep
inelastic scattering (DIS) with polarised leptons on protons allow 
tests of the ${\rm V}-{\rm A}$ structure of
CC interactions to be extended into the high-$Q^2$ regime.
The polarisation dependence of the CC cross sections is
fixed within the Standard Model (SM). Specifically, the SM
predicts, from the absence of right handed charged currents, that the
CC $e^+p$ ($e^-p$) cross section is proportional to the fraction 
of right handed positrons (left handed electrons) in the beam:
\vspace{-1mm}
\begin{equation}
{\sigma_{\rm CC}^{e^\pm p}}({\mathcal P}_e)=
(1\pm {\mathcal P}_e){\sigma_{\rm CC}^{e^\pm p}}({\mathcal P}_e=0) , \nonumber
\vspace{-1mm}
\end{equation}
where the longitudinal polarisation ${\mathcal P}_e$ is equal to $(N_R-N_L)/(N_R+N_L)$
with $N_R$ ($N_L$) being the number of right (left) handed leptons
in the beam. 
\begin{figure}[ht]
\vspace{-1.1cm}
\centerline{
\epsfxsize=5.5cm\epsfbox{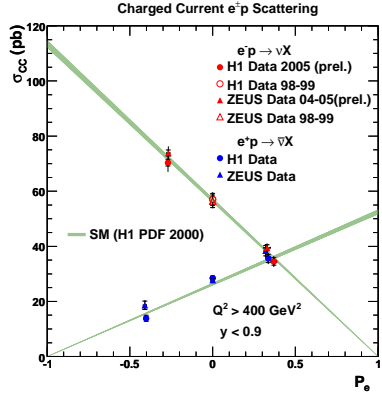}} 
\vspace{-0.4cm}
\caption{
 CC $e^+p$ and $e^-p$ cross sections versus ${\mathcal P}_e$ compared to the SM prediction. 
  \label{ccplot}}
\vspace{-0.51cm}
\end{figure}
H1 and ZEUS published\cite{ccpos} 
the first measurements of the CC $e^+p$ total cross sections 
for positive and negative longitudinal polarisation,
and presented\cite{ccel} the CC $e^-p$ total cross sections.  
The value of the polarisation was typically between $30\%$ and $40\%$.
The measured integrated CC cross sections, 
$\sigma_{\rm CC}^{e^\pm p}$, quoted in the range
$Q^2> 400\,{\rm GeV}^2$ and $y<0.9$, are shown in Fig.~\ref{ccplot}.  
The measurement of the unpolarised
total cross section, in the same phase space, based on data 
collected until $2000$ (HERA-I) is also shown. 
The measurements are compared to SM predictions
based on the H1 PDF $2000$ parametrisation\cite{h1pdf2000}. 
The measurements agree with SM predictions and 
exhibit the expected linear dependence as a function of the polarisation.
Linear fits provide a good description of the data and
their extrapolation to the point ${\mathcal P}_e=-1$ (${\mathcal P}_e=1$) 
yield a fully left (right) handed CC cross section 
for $e^+p$ ($e^-p$) interactions which is consistent with the
vanishing SM prediction. The corresponding
upper limits on the total CC cross sections
exclude the existence of charged currents involving
right handed fermions mediated by a boson of mass below 
$180-208\,{\rm GeV}$ at $95\%$\,confidence level,
assuming SM couplings and a massless right handed $\nu_e$.


\begin{figure}[ht]
\vspace{-0.7cm}
\centerline{\epsfxsize=5.7cm\epsfbox{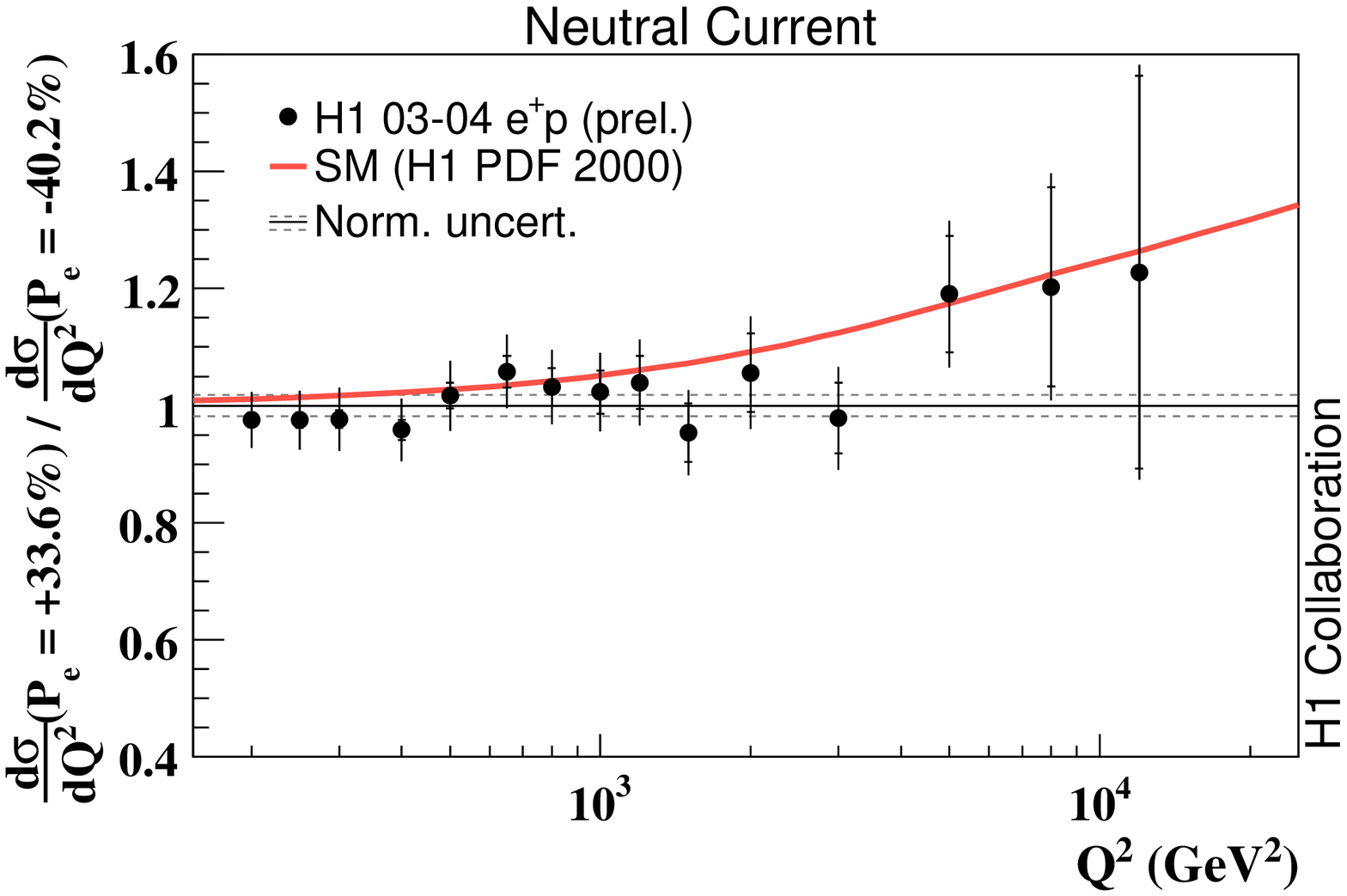}
\epsfxsize=5.7cm\epsfbox{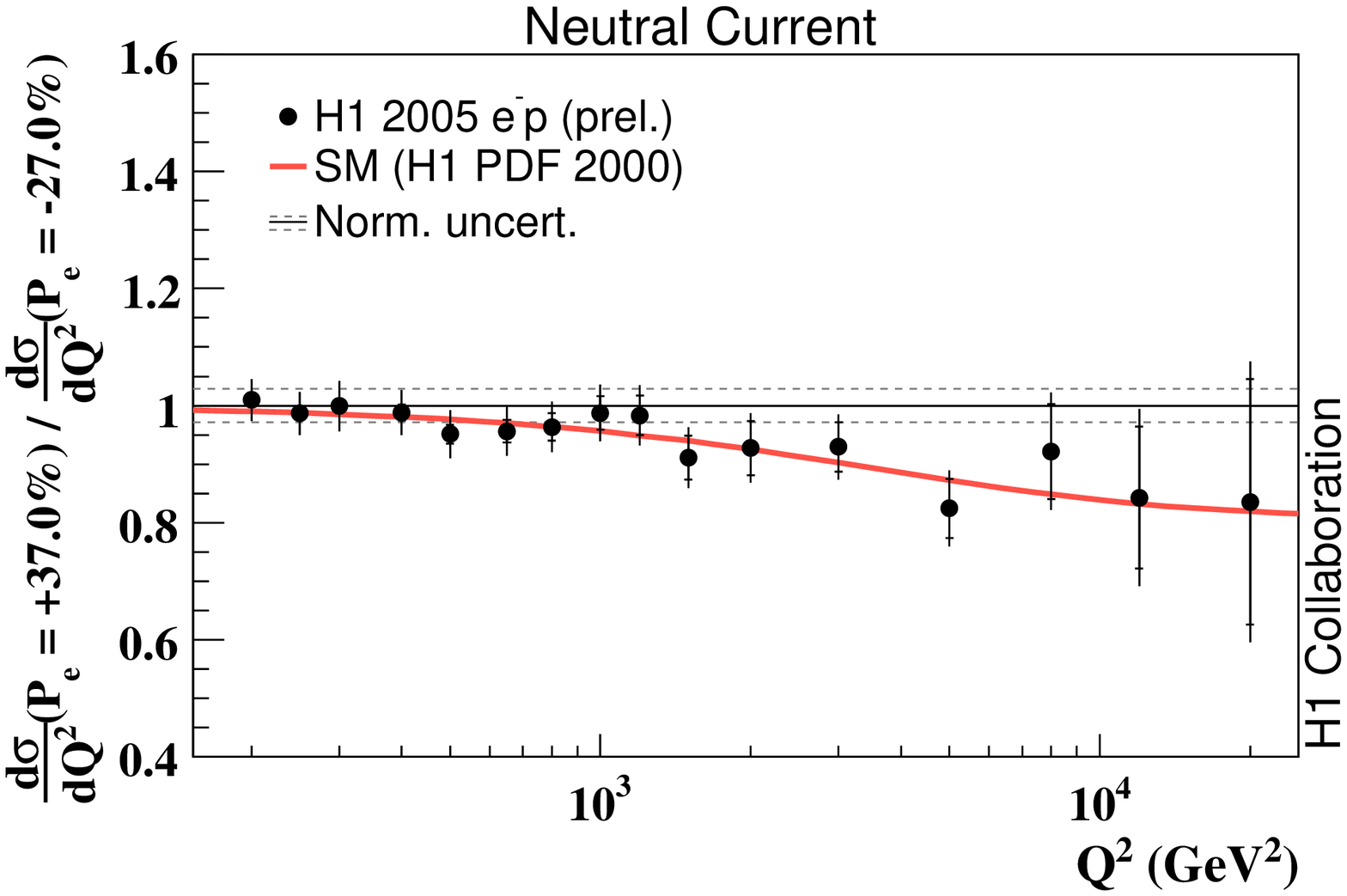}}   
\centerline{\epsfxsize=5.6cm\epsfbox{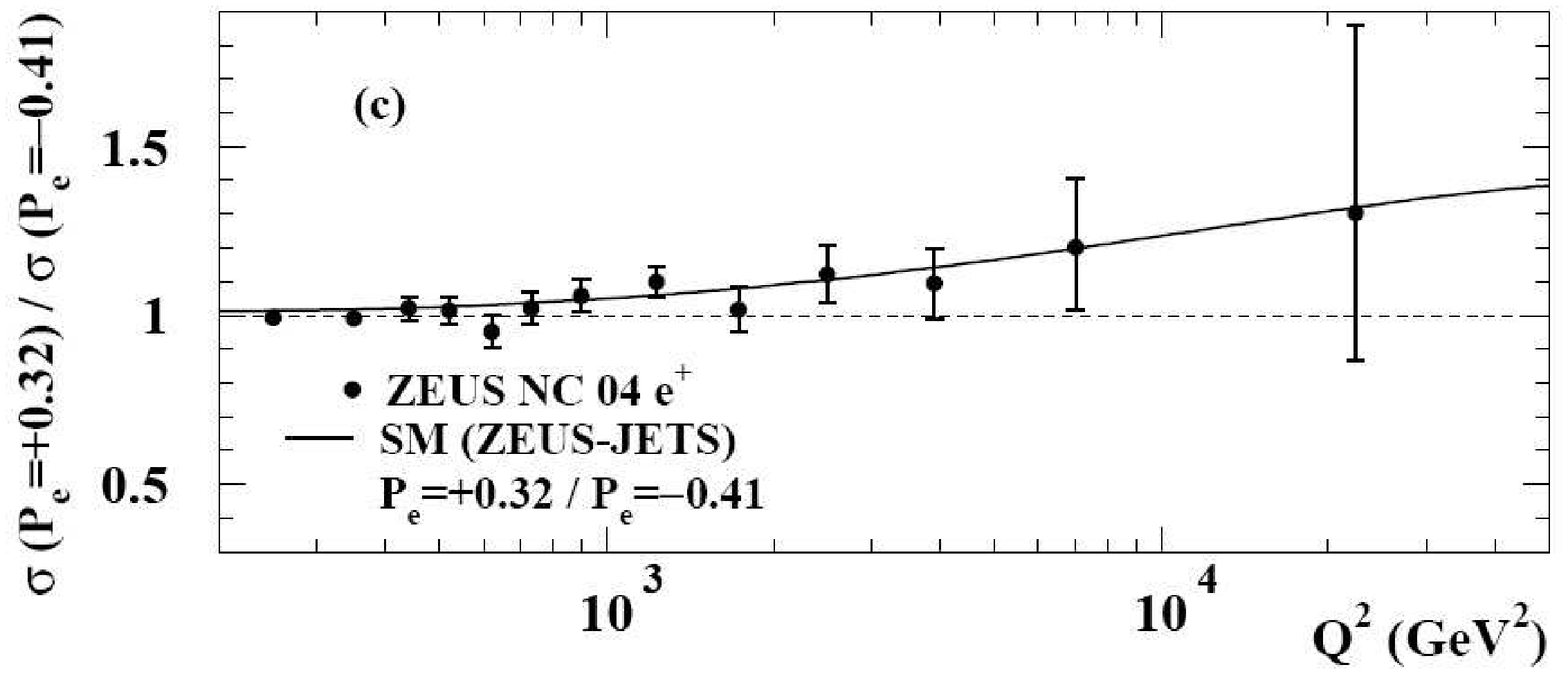}
\epsfxsize=5.6cm\epsfbox{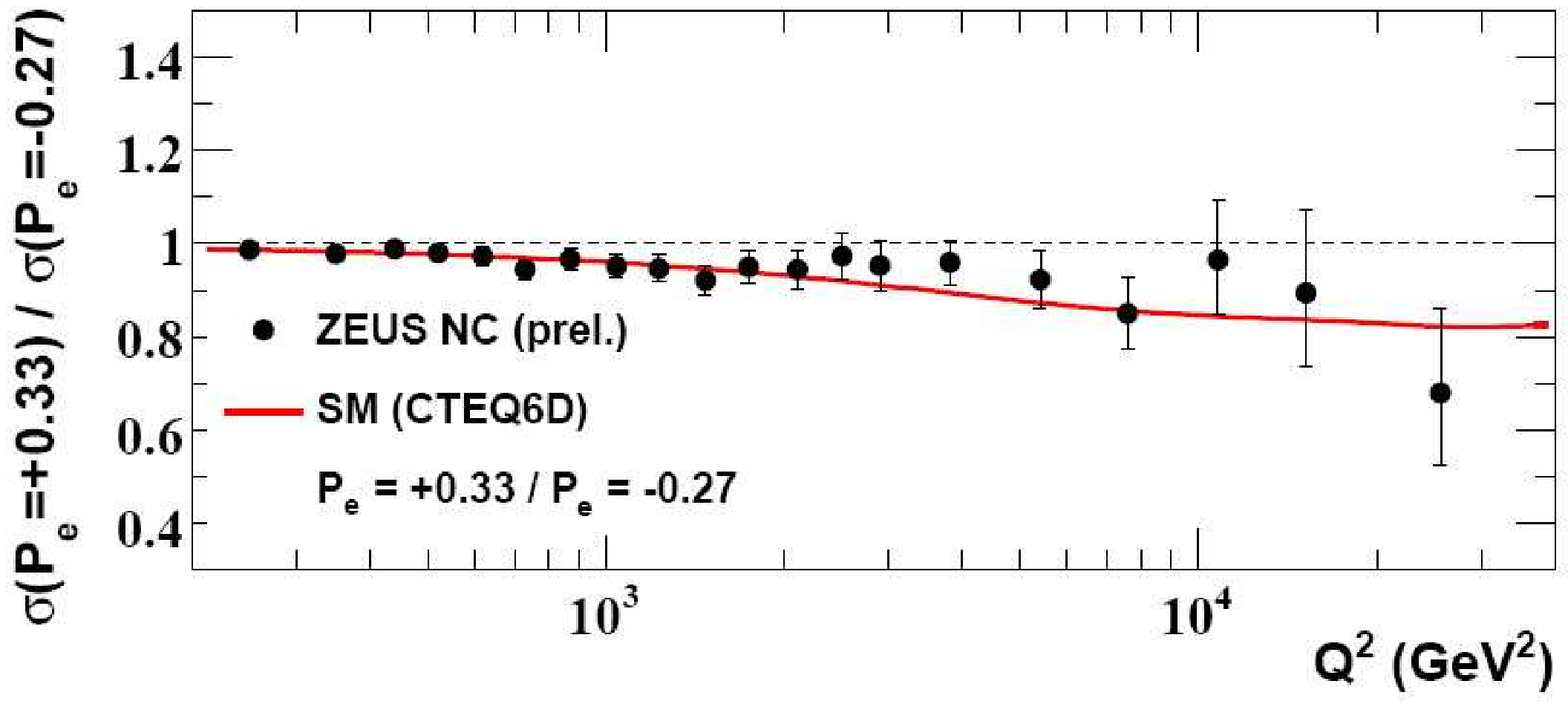}}
\vspace{-0.2cm}   
\caption{
The ratios of the NC cross sections for leptons with positive 
and negative polarisations at HERA-II 
measured by H1 (top) and ZEUS (bottom) as function of $Q^2$,
for $e^+p$ (left) and $e^-p$ (right). 
Solid lines correspond to the SM expectation.
\label{ncplot}}
\vspace{-0.5cm}
\end{figure}


The NC single differential cross sections ${\rm d}\sigma/{\rm d}Q^2$ 
have been measured at HERA-II for $e^+p$ and $e^-p$ scattering
with positively and negatively polarised lepton beams\cite{nc}.
The SM predicts a difference in the cross section for
leptons with different helicity states arising from the chiral
structure of the neutral electroweak exchange. In
Fig.~\ref{ncplot} the ratio of cross sections for positively and negatively
polarised lepton beams is shown separately for $e^+p$ (left plots)
and $e^-p$ (right plots) scattering data. In both cases the ratio is found 
to be consistent, within experimental uncertainties, with unity at low $Q^2$,
indicating little dependence of the cross section on beam polarisation. 
The normalisation uncertainties of the measurements
are not included in the errors bars, but are indicated by the dashed
lines on the upper figures. At higher $Q^2$, the data have a tendency to
deviate from unity. For $e^+p$ scattering the data indicate that
right-handed positrons yield a larger cross section than left-handed ones,
whereas for $e^-p$ scattering the data show the
opposite behaviour. This is consistent with the SM
expectation shown as the solid curve in Fig.~\ref{ncplot}.

\label{subsec:ewfit}

\begin{figure}[Ht]
\vspace{-0.6cm}
\hspace{-60mm}
\centerline{\epsfxsize=6cm\epsfbox{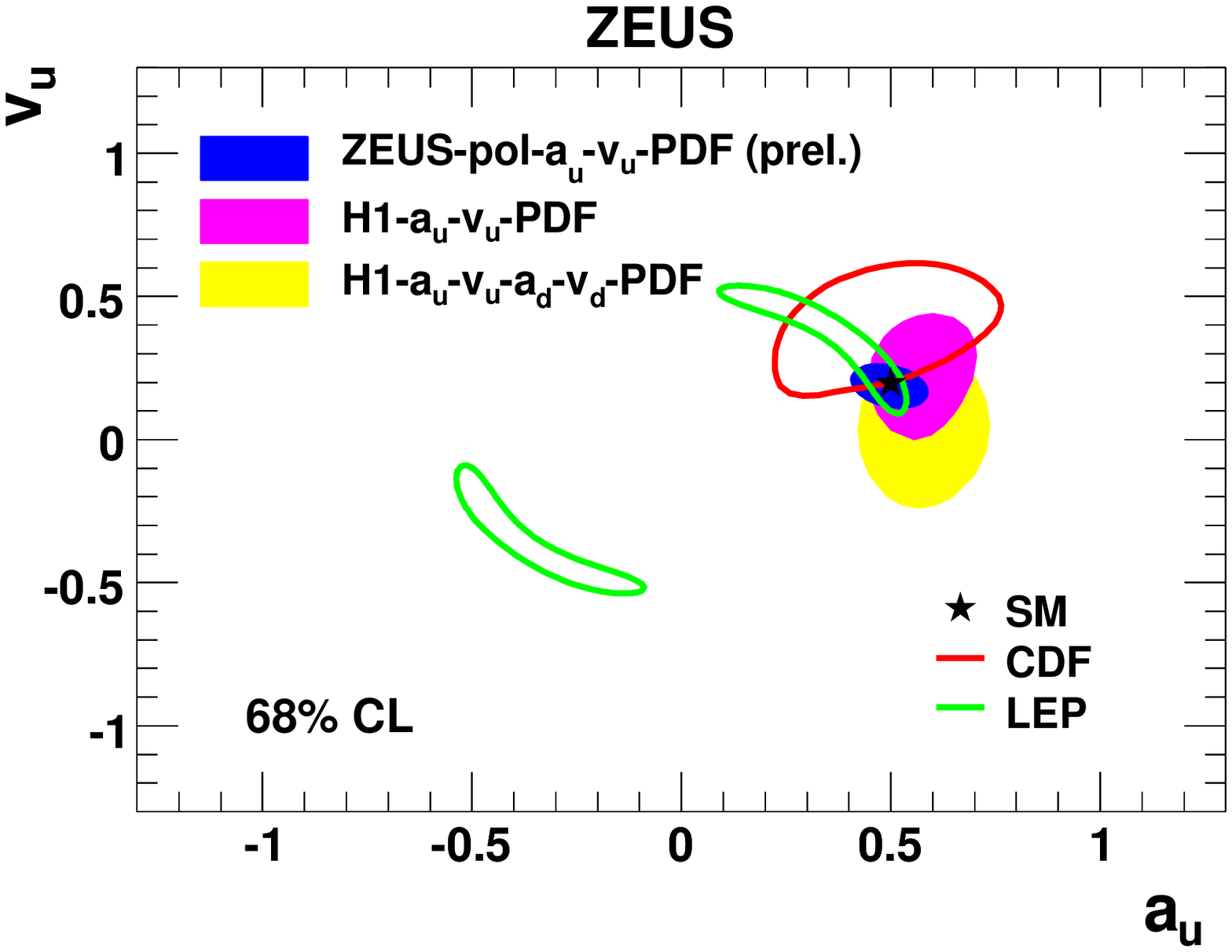}}
\vspace{-44.5mm}
\hspace{500mm}
\rightline{\epsfxsize=6cm\epsfbox{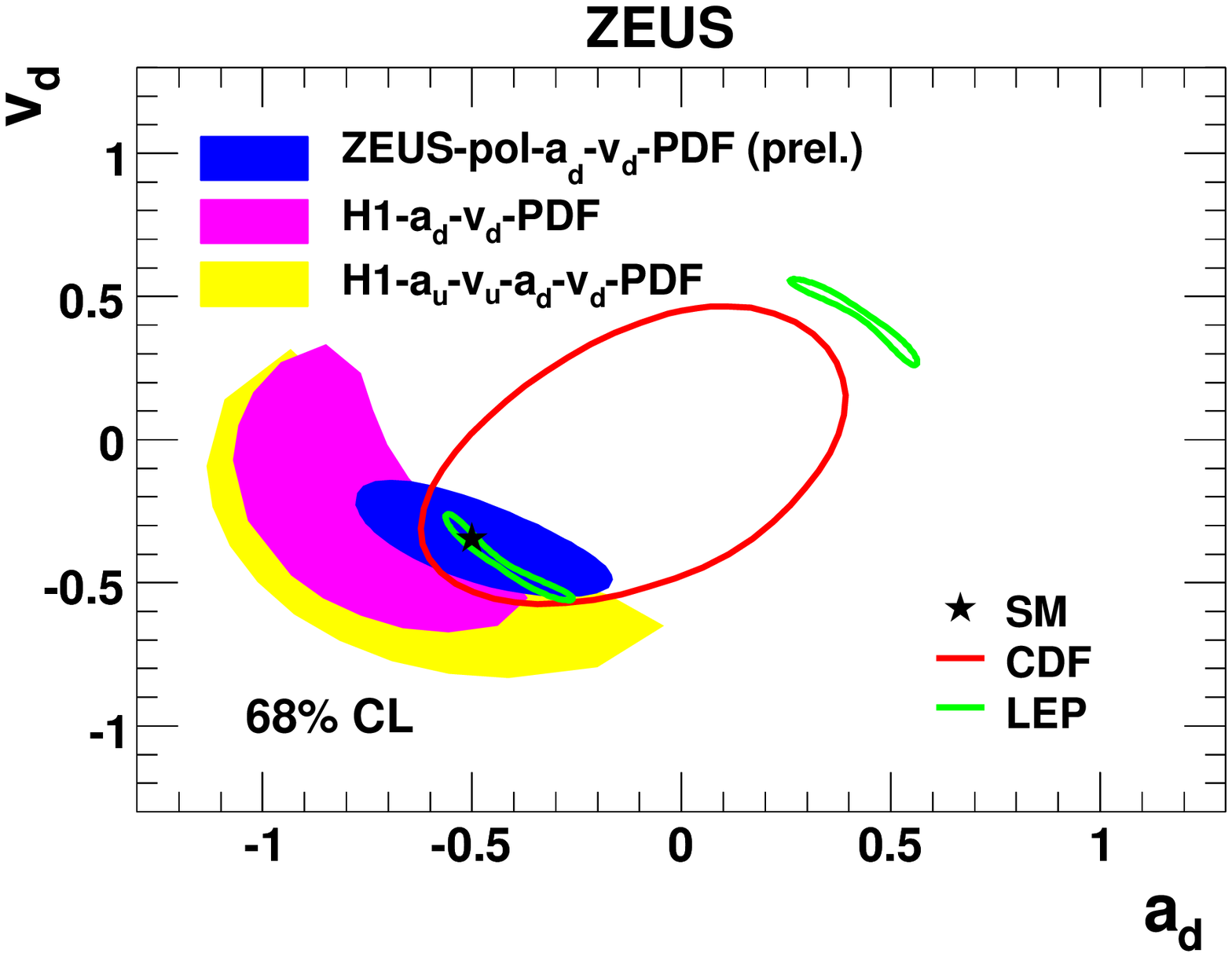}}
\vspace{-0.8cm}   
\caption{
  Contours of the $68\%$ confidence level (CL) on the weak neutral current
  couplings of $u$ (left plot) and $d$ (right plot) quarks to the $Z^0$
  boson.
  \label{ewplot}}
\vspace{-0.7cm}
\end{figure}

At HERA, the NC interactions at high $Q^2$ receive contributions from
$\gamma Z$ interference and $Z^0$ exchange. Thus, the NC data
can be used to extract the weak couplings of 
the light $u$ and $d$ quarks to the $Z^0$ boson.
The CC cross section data help
disentangle the up and down quark distributions and allow a determination of  
a propagator mass $M_{\rm prop}$ from the $Q^2$ dependence of the cross sections.
$M_{\rm prop}$ measured at HERA in the space-like region can be compared with 
direct $W$ boson mass measurements obtained in the time-like region
by the Tevatron and LEP experiments.
Combined electroweak and QCD fits at HERA have been 
performed\cite{h1ewfit,ewfit}, taking into account the correlation between 
the electroweak parameters and the parton distribution functions (PDFs).
The ZEUS collaboration extend their NLO QCD fit\cite{zeuspdf} 
to inclusive NC and CC data, inclusive jet data in DIS and dijets 
in photoproduction, to determine simultaneously the PDFs and electroweak
parameters. This fit also includes, for the first time,  the $e^{\pm}p$ 
polarised NC and CC double-differential HERA-II cross sections.
The resulting PDFs are very similar to those obtained from the fit to  
HERA-I data, with slightly improved uncertainties on the $u$-quark at high $x$,  
due to additional data. In this analysis, the fits have been performed 
by fixing either the $u$ or the $d$ quark couplings to their SM values
(fits $v_d$-$a_d$-${\rm PDF}$ and $v_u$-$a_u$-${\rm PDF}$).
The resulting one-sigma contours are shown in Fig.~\ref{ewplot}.
The results are consistent with the electroweak Standard Model and
the precision is better for the $u$ quark as expected.
Comparison to the H1 fits to unpolarised HERA-I data
shows that, while the uncertainty on the axial-vector 
couplings stays about the same, the precision of the determination of the
vector couplings is improved by a factor of 2--3 due to
additional sensitivity of the polarised NC data.
Figure~\ref{ewplot} also shows the results of the H1 fits
in which the vector and axial-vector couplings of $u$ and $d$ 
quarks are fitted simultaneously 
(fits $v_u$-$a_u$-$v_d$-$a_d$-${\rm PDF}$) and
similar results obtained recently by the CDF experiment and at LEP.
The HERA determinations have comparable precision to that from the Tevatron
and resolve any sign ambiguity and the ambiguities between $v_u$ and 
$a_u$ of the determinations
based on observables measured at the $Z^0$ resonance.
Exploiting the $Q^2$ dependence of the charged current data,
the propagator mass has been measured to be
$M_{\rm prop}=82.8\pm 1.5\pm 1.3 {\rm ~GeV}$, 
which is in agreement with the direct measurements of the $W$ boson mass. 

\begin{figure}[Ht]
\vspace{-0.7cm}
\centerline{
\epsfxsize=5.8cm\epsfbox{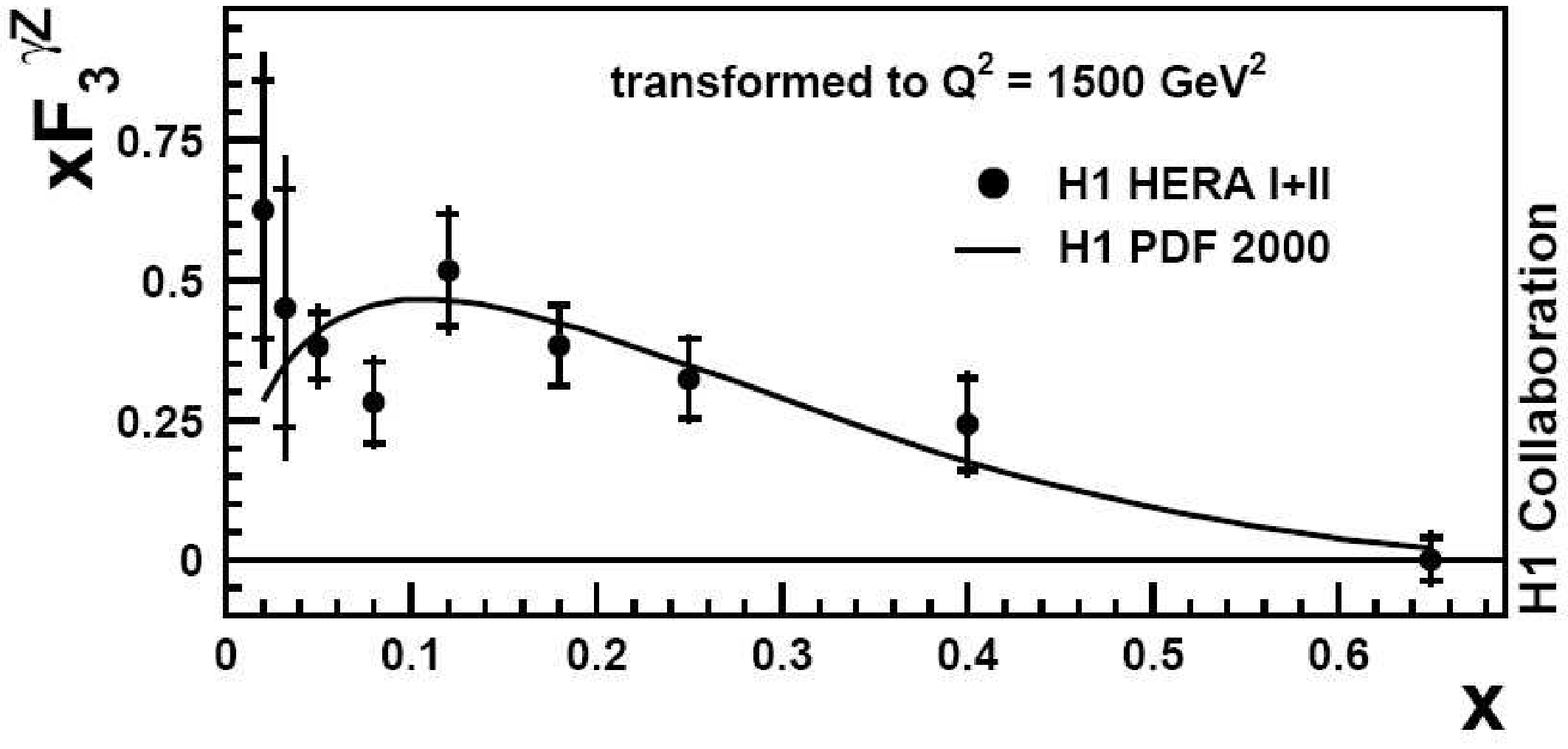}
\epsfxsize=5.8cm\epsfbox{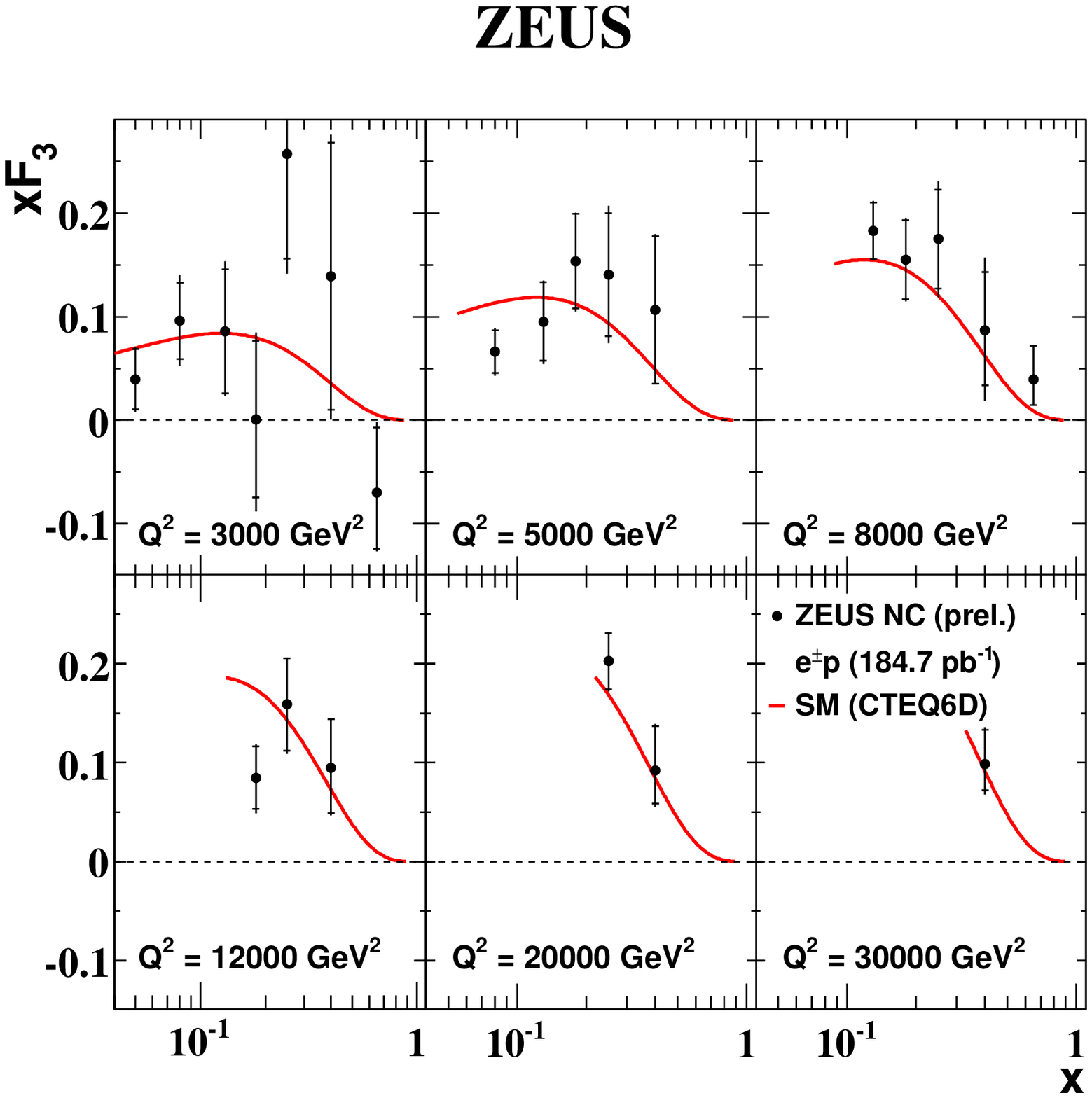}}   
\vspace{-60mm}
\hspace{-30.8mm}
\centerline{\epsfxsize=5.35cm\epsfbox{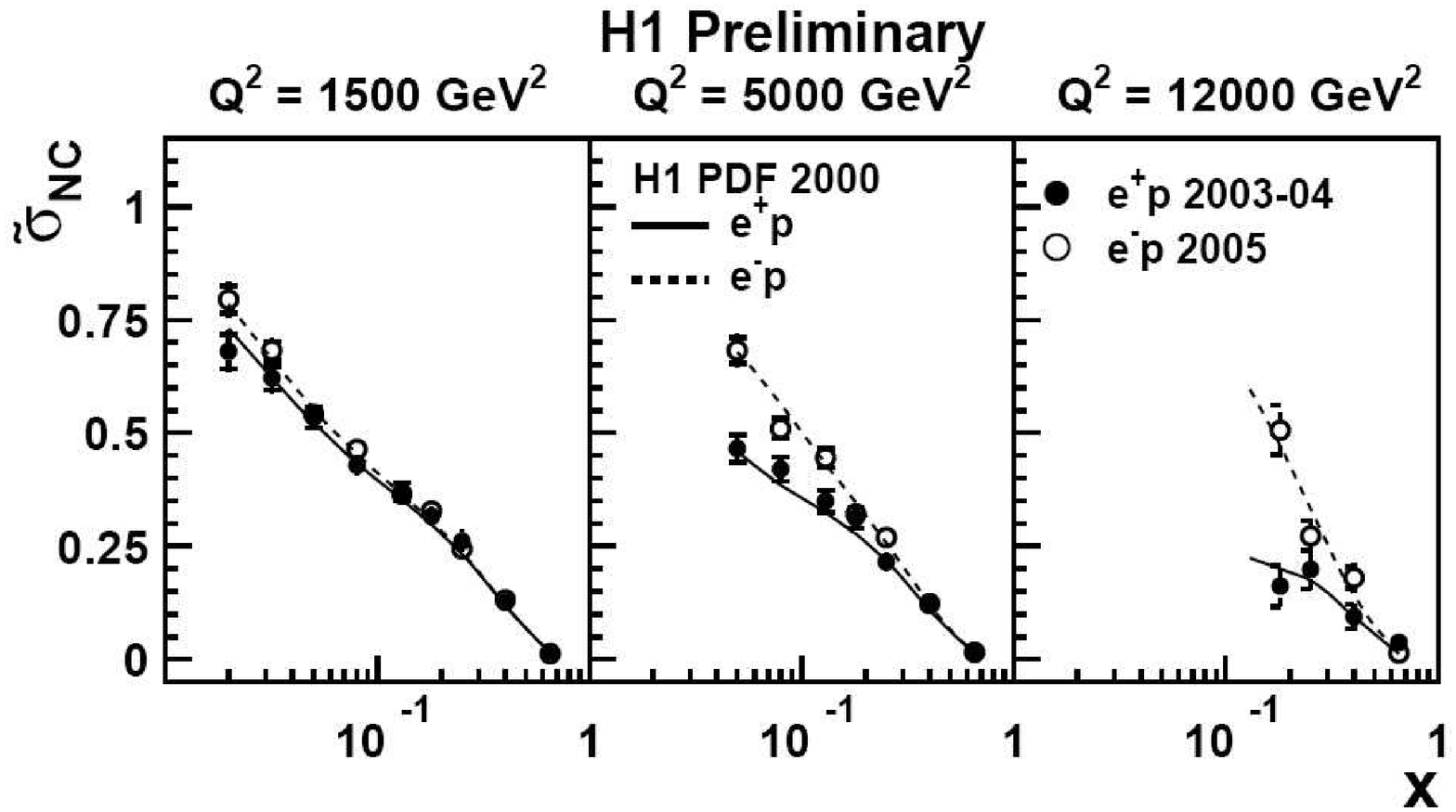}}
\vspace{30mm}
\vspace{-0.5cm}
\caption{The NC reduced cross sections 
 $\tilde{\sigma}^\pm_{NC}(x,Q^2)$ measured by H1 at HERA-II (left top), 
 the structure functions $xF_3$ from ZEUS (right) and
 the averaged H1 structure function $xF_3^{\gamma Z}$ for $Q^2
 =1500{\rm\,GeV}^2$ (left bottom) compared to 
 QCD expectations.
\label{xf3plot}}
\vspace{-0.4cm}
\end{figure}

The amount of data collected at HERA-II is already greater than that of HERA-I.
In particular, a significant increase of integrated luminosity is achieved in 
the $e^-p$ mode, from $\approx$ 15 pb$^{-1}$ to more than 100 pb$^{-1}$. 
At HERA-I the statistics of the $e^-p$ data
was a limiting factor for the precision of the $xF_3$ determination.
Profiting from the enlarged statistics and reduced systematic
uncertainties, the previous measurement of $xF_3$
has been updated\cite{nc} using
HERA-II 2003-2005 $e^{\pm}p$ NC cross section data at high~$Q^2$.
Fig.~\ref{xf3plot} (left top) shows the comparison of the unpolarised
$e^-p$ and $e^+p$ HERA-II H1 data for three different $Q^2$ values.  
At high $Q^2$, the NC cross section in $e^-p$ scattering is significantly 
larger than that in $e^+p$ scattering due to the different sign of the
$xF_3$ contribution to the cross section for different leptons
polarities.  $xF_3$ determined by ZEUS from the difference of 
$e^-p$ and $e^+p$ cross sections is shown in Fig.~\ref{xf3plot} (right).
The dominant contribution to $xF_3$ arises from the $\gamma Z$
interference, which allows $xF^{\gamma Z}_3$ to be extracted according to
$xF^{\gamma Z}_3\simeq -xF_3(Q^2+M^2_Z)/(a_e\kappa Q^2)$ by neglecting
the pure $Z$ exchange contribution, which is suppressed by the small vector
coupling $v_e$. 
Here, $\kappa^{-1}=4M_W^2/M_Z^2(1-M_W^2/M_Z^2)$,
$M_Z$, $M_W$ are the vector boson masses and
$a_e$ is the axial-vector coupling of the electron.
This structure function is non-singlet and has little dependence on $Q^2$.
The measured $xF_3^{\gamma Z}$ at different $Q^2$ values can thus be averaged
taking into account the small $Q^2$ dependence. 
The averaged $xF_3^{\gamma Z}$,
determined by H1 for $Q^2 =1500{\rm\,GeV}^2$, is shown
in Fig.~\ref{xf3plot} (left bottom) compared to
the QCD fit result.
The structure function $xF^{\gamma Z}_3$ determines both the shape and
magnitude of the valence quark distributions independently of
the sea quark distributions. 

\vspace{-0.5cm}

\section{FNAL, JLAB, RHIC and new methods for data analysis}
\vspace{-0.1cm}
In this section, we summarise a number of new results from 
experiments at Fermilab, JLab and RHIC, and discuss several 
new approaches for determining PDFs and their uncertainties.


Recently, the importance of reducing the proton PDF uncertainties, 
especially at high $x$, has become increasingly apparent, 
both for precision SM studies and new physics searches at hadron colliders. 
The CDF and D0 collaborations presented a number of QCD and electroweak measurements 
that could provide additional constraints on the PDFs\cite{cigdem,mikko}. 
Of particular importance is the, poorly known, high-$x$ gluon distribution, which can be constrained 
using hadron collider jet data.   
CDF and D0 presented new measurements of inclusive jets\cite{cigdem,mikko} in $p\bar{p}$ collisions.  
The data agree well with NLO QCD predictions over the measured $E_T$ range, which extends to $\sim 600-700$ GeV. 
The precision of the data, which is limited by the jet energy scale uncertainty, 
is at a level where these measurements should 
provide significant additional constraints on the high-$x$ gluon. In particular, 
in the most forward region measured by CDF ($1.6 < |y| < 2.1$) experimental 
uncertainties are smaller than those from the PDFs. Note that 
measurements over a wide range of rapidity are important to discriminate 
between PDF effects and new physics signals. 
CDF and D0 also presented a number of other results\cite{cigdem,mikko}, including 
measurements of the forward-backward charged lepton asymmetry from 
$W^\pm \rightarrow l^\pm \nu$ decays, which constrains the $d/u$ ratio, 
a new measurement of the forward $W \rightarrow e\nu$ cross section 
and studies of $Z \rightarrow \mu\mu$, $Z \rightarrow \tau_e \tau_h$, $Z+jets$ and dijet azimuthal decorrelations.

NLO QCD calculations of jet cross sections are CPU intensive, rendering their use 
in QCD fits challenging. Global fit groups\cite{cteq,MRST04} use LO cross sections + $k$-factors 
to approximate the NLO result. At this workshop, an approach was presented\cite{gridnlo}, 
in which an NLO QCD program is used to calculate a grid of weights in $(x,Q^2)$, 
which can be convoluted with any PDF to give fast NLO 
predictions\footnote{A similar grid technique was also 
presented\cite{wobisch} in the Hadronic Final States session.}. The 
accuracy of the grid-computed cross sections is improved, without significant 
cost to CPU, by using coordinate transformations and high order interpolation 
between grid points. This technique will allow the rigorous inclusion 
of HERA, Tevatron and LHC jet data in future QCD fits.

\begin{figure}[Htp]
\vspace{-1.3cm}
\centerline{\epsfxsize=7.8cm\epsfbox{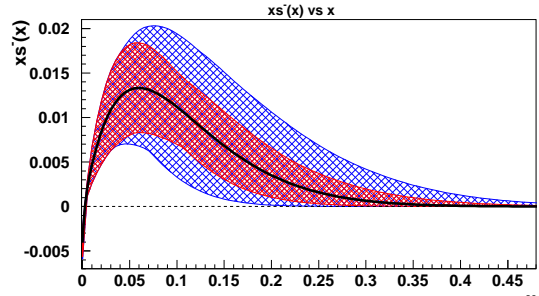}}
\vspace{-0.5cm}
\caption{The distribution of $xs^-$ at $Q^2 = 16$ ${\rm GeV}^2$.\label{fig-mason}}
\vspace{-0.4cm}
\end{figure}
NuTeV presented the final measurement of the 
difference between the strange and antistrange quark distributions\cite{mason}. 
This is a topic of particular interest, since it has been speculated\cite{sa1} that 
a non-zero difference -- or strange asymmetry -- could explain the almost $3\sigma$ 
difference between the NuTeV $\sin^2\theta_W$ result\cite{sin2thw} and the world average. 
NuTeV have performed the first complete NLO QCD analysis of CC $\nu N$ and $\bar{\nu} N$ scattering, 
with two oppositely charged muons in the final state, giving direct access to the 
strange quark content of the nucleon. Figure~\ref{fig-mason} shows the distribution 
of $xs^-(x) = xs(x)-x\bar{s}(x)$ extracted from the fit, which is positive at 
moderate $x$, such that $S^- \equiv \int_0^1 xs^-(x) dx = +0.00196 \pm 0.00143$. 
NuTeV are now updating their $\sin^2\theta_W$ 
result, which originally assumed $S^-=0$, in light of this result, 
and also taking into account other recent measurements, such as the BNL $K_{e3}^+$ measurement\cite{k+e3}. 
Note also, that the data prefer an asymmetry which forces $s^-(x)$ negative at low values of 
$x_0 \lesssim 0.004$, in contrast to some theoretical models, which suggest 
larger values of $x_0$. 

The E03-103 experiment at JLab have performed a new measurement of the EMC effect in light nuclei\cite{seely}. 
The characteristic $x$-dependence of the ratio $\sigma_A/\sigma_d$ 
(i.e. the so-called ``EMC effect''), has been well studied. 
While broad features can be explained, no single model has successfully described the effect over all $x$. 
To date, much of the experimental effort has been concentrated on heavy nuclei ($A > 4$).  
However, calculations of the EMC effect in $^3$He and $^4$He predict large differences 
in both magnitude and $x$-dependence.  
E03-103 have measured inclusive electron scattering, using a 5.8 GeV beam, providing the first measurement of the EMC effect in $^3$He for $x > 0.4$, 
and improving upon existing measurements\cite{emcprevious} of $^4$He. 
The results indicate that the effect 
in $^4$He is similar to that in carbon and the effect in $^3$He could be larger than expected, 
although the strength of the latter conclusion is limited by large uncertainties in proton excess corrections. 

JLab also presented results on the phenomenon of parton-hadron duality in semi-inclusive scattering. 
Duality, for example the fact that the average behaviour of nuclear resonances exhibits the scaling behaviour expected 
in pQCD, is well established in inclusive scattering, but has yet to be studied to the same extent in semi-inclusive data. 
The merit of semi-inclusive processes lies in the ability to identify 
individual quark species in the nucleon by tagging specific mesons in the final state. 
The E00-108 experiment at JLab presented\cite{ent}  
preliminary measurements of semi-inclusive pion electroproduction, $eN \rightarrow e' \pi^\pm N$, 
using a 5.5 GeV electron beam on proton and deuteron targets, 
for $1.8 < Q^2 < 6.0$ ${\rm GeV}^2$, $0.3 \le x \le 0.55$, and elasticities, $z = E_\pi / \nu$, in the range $0.35-1$.   
The results indicate that the ratio of the unfavoured to favoured fragmentation functions, 
$\frac{D^-}{D^+} = \frac{4-{N_\pi^+}/{N_\pi^-}}{4{N_\pi^+}/{N_\pi^-}-1}$, shows no dependence on $x$   
and a smooth slope as a function of $z$, in accordance with the expectations of the onset of duality. 
It was argued that these data are also consistent with the idea that pion electroproduction 
is related to low energy factorisation. If so, future semi-inclusive measurements could 
give excellent access to the flavour structure of the nucleon.

\begin{figure}[Htp]
\vspace{-0.9cm}
\centerline{\epsfxsize=5.4cm\epsfbox{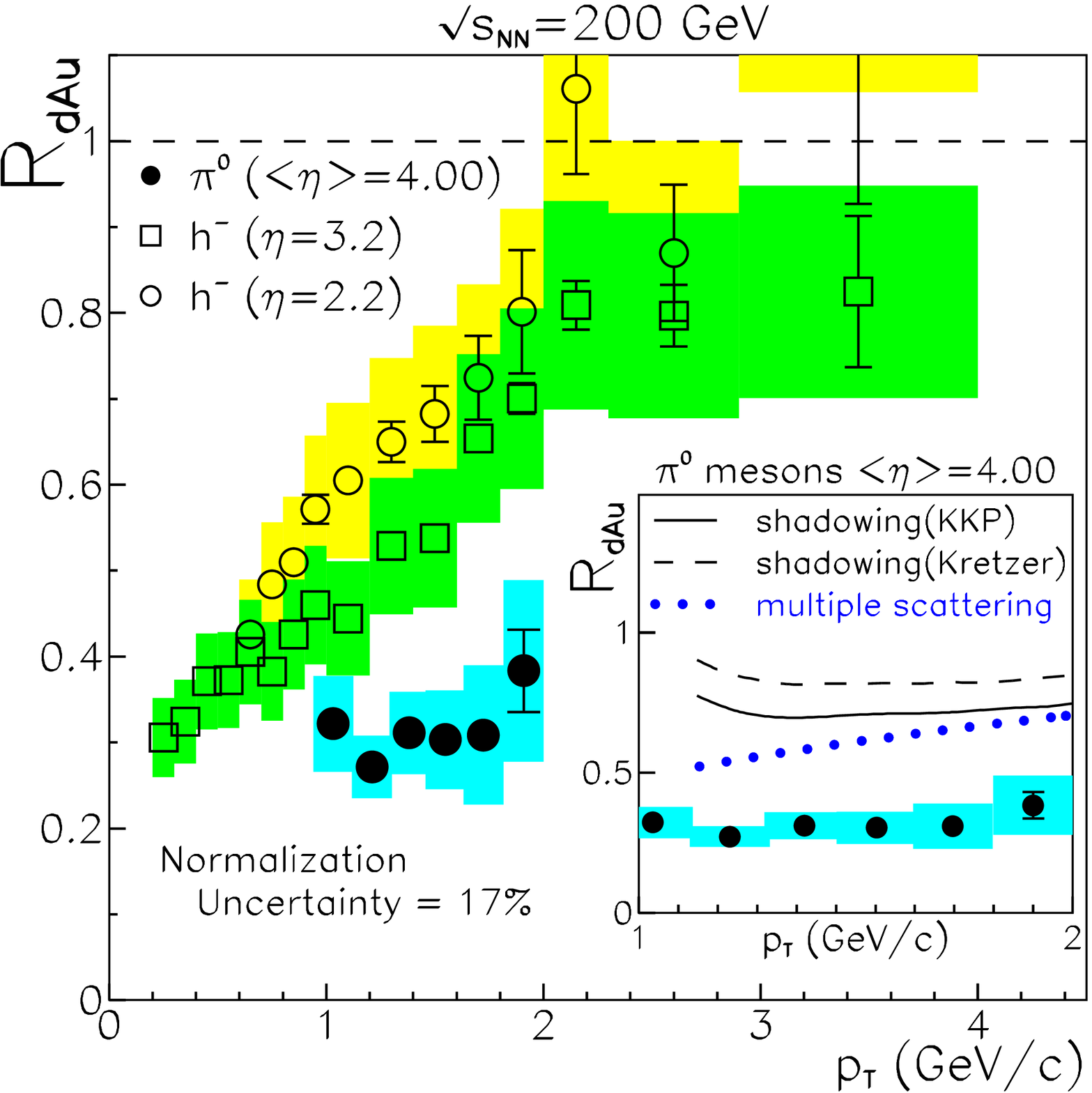}\hspace{0.2cm}
\epsfxsize=5.4cm\epsfbox{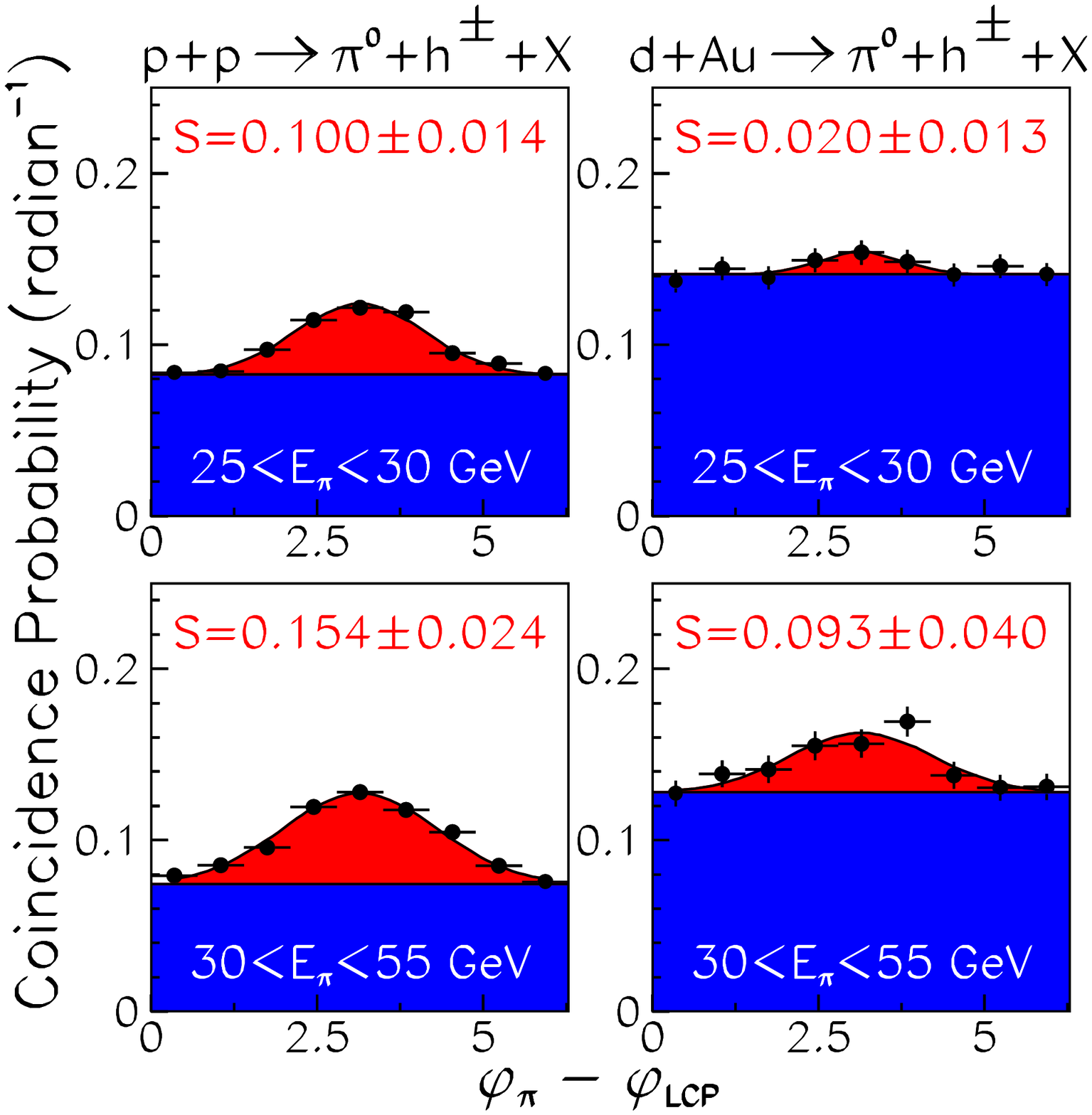}}
\vspace{-0.2cm}
\caption{Left: nuclear modification factor, $R_{\rm dAu}=\frac{1}{2\times 197}{\sigma_{\rm d+Au}}/{\sigma_{\rm p+p}}$, as a function of $p_{\rm T}$ for $\pi^0$ 
mesons at $\langle \eta \rangle = 4.00$ and charged hadrons at smaller pseudorapidities. Error bars are statistical, while the shaded boxes are point-to-point systematics. 
Right: azimuthal correlations between forward $\pi^0$ and charged hadrons at mid-rapidity. 
\label{fig-gagliardi}}
\vspace{-0.5cm}
\end{figure}
The STAR collaboration have searched for parton saturation effects at RHIC\cite{star}. 
The term saturation describes the idea that at very small $x$, the gluon density is so large, 
that gluons begin to recombine, leading to a slowing of the growth and giving rise to non-linear 
terms in the evolution equations. 
The state of saturation 
is often referred to as the Colour Glass Condensate (CGC). Nuclear environments, such as 
RHIC, are ideal places to search for signs of saturation, due to the $A^{1/3}$ enhancement 
of the parton densities in nuclei, relative to the proton. STAR presented results on forward $\pi^0$ 
production in p+p and d+Au collisions, in the range $3.0 \lesssim \eta \lesssim 4.2$, 
at $\sqrt{s} = 200$ GeV. Measurements at forward rapidities ensure that low values of gluon-$x$ are probed. 
The p+p yields generally 
agree with NLO pQCD predictions, while the d+Au yield is suppressed at forward rapidities, 
as shown in Fig.~\ref{fig-gagliardi} (left). 
It was argued that the $p_{\rm T}$ dependence 
of the d+Au yield is consistent with a model treating the Au nucleus as a CGC for forward 
pion production\cite{star}. The azimuthal correlations of the forward $\pi^0$ with charged hadrons at 
mid-rapidities (Fig.~\ref{fig-gagliardi} (right)) show a recoil peak in p+p that is 
suppressed in d+Au collisions at low $E_\pi$; in qualitative 
agreement with a gluon saturation picture of the Au nucleus. Note, however, that the PHENIX experiment at RHIC do 
not see significant differences in the p+p and d+Au azimuthal correlations\cite{phenix}. STAR 
will soon make further measurements, utilising a new forward meson spectrometer, that could 
elucidate the source of the observed suppression.

Several presentations were made in which alternative approaches to PDF 
determination were explored.  The NNPDF collaboration presented the latest results 
on a neural network (NN) approach to PDF fitting\cite{nnpdf}. This approach attempts to avoid any bias introduced 
from a choice of functional form for the PDFs and should provide more faithful estimates of the PDF  
uncertainties. The NNPDF best-fit of the non-singlet structure function, $F_2^{NS}(x,Q^2)=F_2^p(x,Q^2)-F_2^d(x,Q^2)$,  
agrees with the predictions of other PDF sets, and gives a better description 
at large $x$, as well as larger error bands in the region where there is no data. 
Work is now in progress to construct a full set of PDFs. 
A related approach was presented by the 
SOMPDF group\cite{sompdf}. This technique is based on a specific class of NN, 
called the Self Organising Map (SOM). It was argued that SOMs allow better control 
of systematic bias, by replacing the fully automated procedure of standard NNs, 
with an interactive fitting procedure. 
Finally, the prospects for determining PDFs and their uncertainties 
using a Bayesian statistical approach, was discussed\cite{cowan}. This work is at an early stage, 
but it will be extremely interesting to see the results when they become available.


\vspace{-0.4cm}

\section{Theory}

\vspace{-0.0cm}

The main topics were updates in determinations 
of proton and nuclear partons distributions 
and the ever increasing sophistication (complication?) of theory.
The former included the implementation of new heavy flavour prescriptions 
and/or NNLO corrections and new data in fits. 
The latter included recent developments in small-$x$ resummations,
with hopefully the beginning of detailed understanding and 
phenomenology. Matters are 
getting  more involved in the very small $x$ region with 
non-linear effects, and the saturation scale seems to be becoming ever
smaller. 

Starting with fits to partons distributions, Guffanti 
presented a NNLO analysis of
non-singlet parton distributions\cite{Guffanti} $u_v(x,Q^2)$ and $d_v(x,Q^2)$
by fitting to $F^{p,d}_2(,x,Q^2)$, for $x>0.3$ and $F_2^p-F^d_2$. 
The result is $\alpha_S(M_Z^2)=0.1134 
\pm 0.0020$, in good agreement with other NNLO fits to only structure function 
data. Alekhin made his fits more global\cite{Alekhin} by 
including E605 Drell-Yan data\cite{E605} and E866
Drell-Yan ratio data\cite{E866}. 
The fit has no problems with this data, and it improves the 
accuracy on the high-$x$ sea and 
gives his first real constraint on $\bar u - \bar d$.
In fact using $\Delta \chi^2=1$ the sea uncertainties are very small.
The theoretical input to the DIS fit is
the massless NNLO QCD corrections for the light quarks and 
gluons (splitting\cite{NNLOs} and coefficient functions\cite{CF,NNLODY};
account of the heavy quarks up to $O(\alpha_S^2)$\cite{nlocalc} in FFNS;
account of the target-mass corrections, Fermi-motion in deuterium, and 
twist 4 terms; 
and the massless $O(\alpha_S^3)$ corrections\cite{NNNLO} to the 
coefficient functions.  
There is disagreement with the definition of NNLO regarding heavy quarks 
-- this is conventionally only NLO corrections in the FFNS scheme. 
The last step is only part of a full N$^3$LO correction -- lacking the 
splitting functions and 
not necessarily indicative except at high $x$.
Reasonable stability is claimed down to $Q^2=0.5\GeV^2$ for 
$0.06\leq x \leq 0.12$ -- perhaps possible since most corrections are 
small in this region.

There are some issues with new data in the fits. Most interesting
was the NuTeV structure 
function data\cite{NuTeVsf} which is larger than the similar 
CCFR data\cite{CCFR} at high $x$ and is 
useful for flavour separation. Comparison to this data relies on nuclear 
corrections. A determination of 
these was reported by Kumano\cite{Kumano}. 
CTEQ find these NuTeV data difficult to 
fit\cite{Tung}, whilst MRST only obtain a good fit by 
severely reducing the nuclear 
correction\cite{MRSTdis06}, implying that this is 
different for neutrinos than for charged leptons.
The important information in the data is for the 
region $x< 0.3$ which is  not too sensitive 
to the nuclear corrections, but the problem caused much interest. 
Recent CHORUS data\cite{Chorus} are more similar to CCFR data.
There are also some changes in global fit procedures. 
CTEQ\cite{Tung} include all HERA data  and fit directly to
cross-sections for first time which requires $F_L(x,Q^2)$ at high $y$.
They also implement a new heavy flavour prescription\cite{acotchi} 
(see the Heavy
Flavour session summary). It was noted that the new fit overshoots 
the high-$y$ data, with the match between theory and data requiring the use 
of the photoproduction background systematic error at more than 2 $\sigma$,   
i.e. use of the systematic errors can remove the high-$y$ $F_L(x,Q^2)$
turnover. MRST implement a full NNLO VFNS\cite{nnlovfns} and use the  
Drell-Yan cross-sections\cite{NNLODY} for the first time, 
leading to (provisional) full NNLO
partons with uncertainties\cite{MRSTdis06}. 
The improvement in the charm procedure affects the gluon compared to 
the approximate MRST2004 NNLO fit\cite{MRST04}. The quality of the  fit at 
NNLO is consistently better than NLO, but not 
for Drell-Yan data. There is a definite tendency for 
$\alpha_S(M_Z^2)$ to go down with
all the changes -- at NLO $\alpha_S(M_Z^2)=0.121$ and at 
NNLO $\alpha_S(M_Z^2)=0.119$.

\begin{figure}[ht]
\vspace{-2.2cm}
\centerline{\hspace{-0.5cm}\epsfxsize=2in\epsfbox{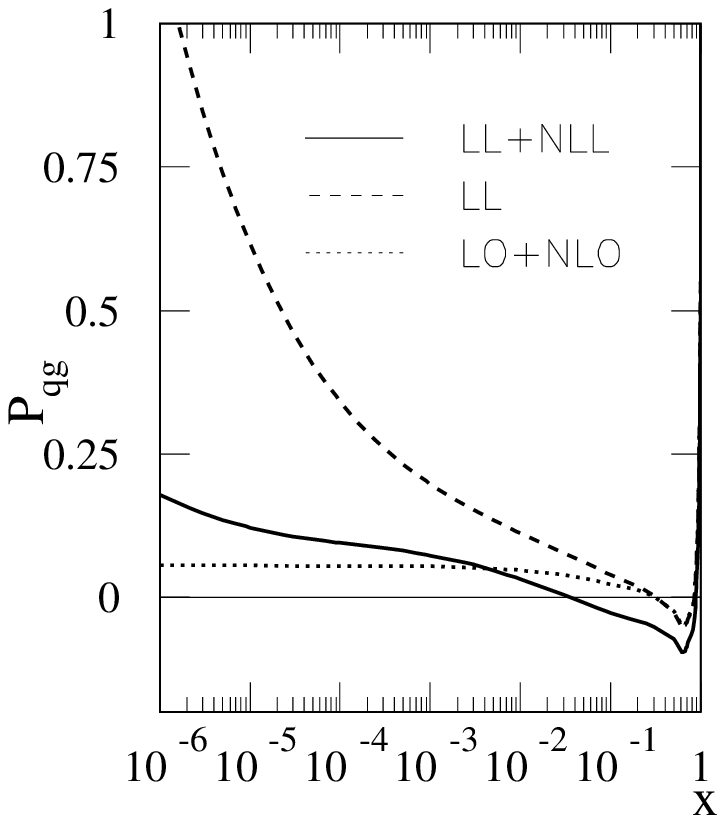}
\epsfxsize=2.07in\epsfbox{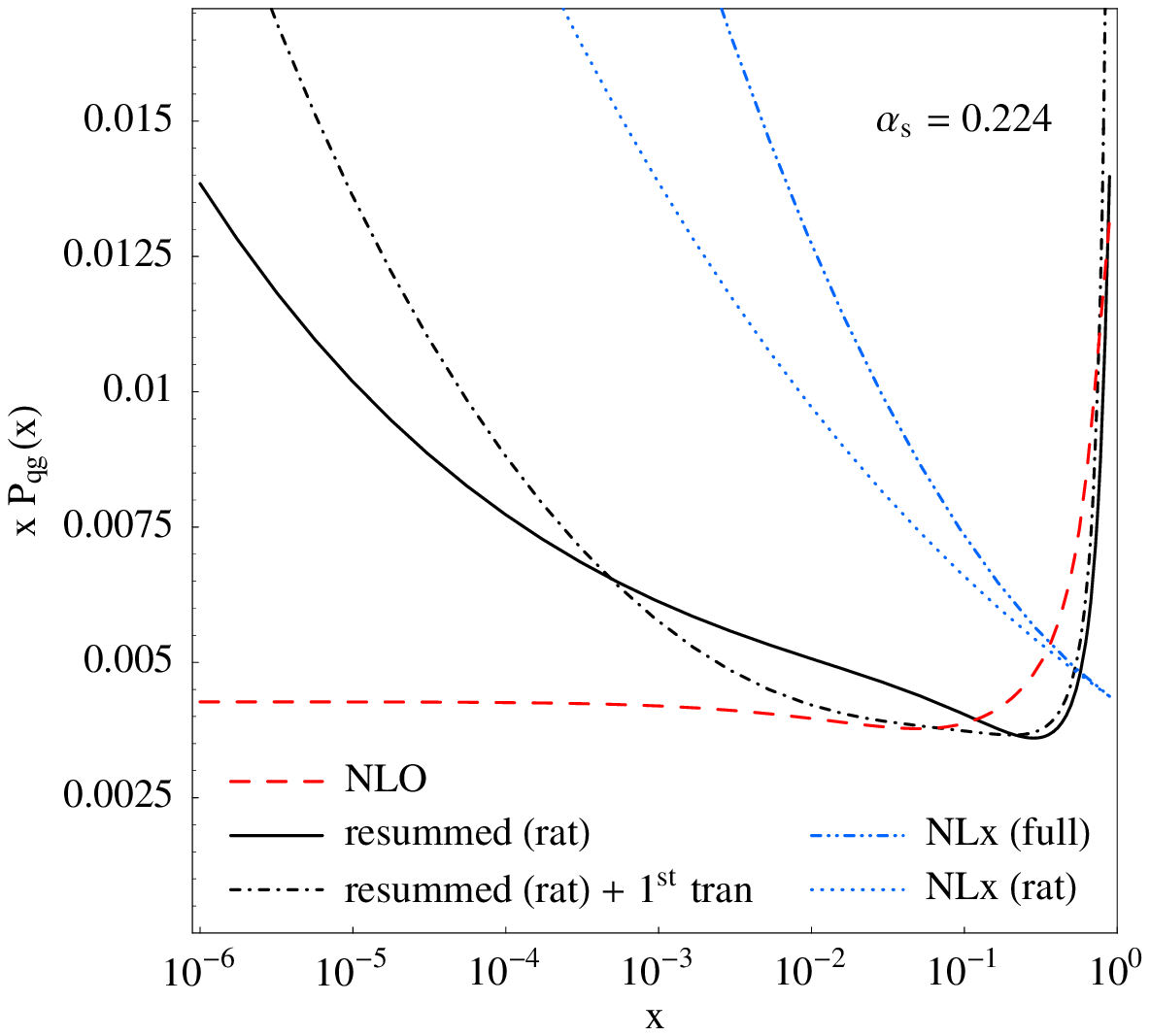}}   
\vspace{-0.4cm}
\caption{The splitting function $P_{qg}$
in two approaches including small $x$ resummations.
  \label{fig1}}
\vspace{-0.7cm}
\end{figure}

There were presentations on various ways to include small-$x$ 
resummations from 
the BFKL equation on top of the fixed order expansion. 
White showed a resummation of $\ln(1/x)$ terms 
along with running coupling corrections\cite{Thorne}, 
which produces mainly analytic results with small numerical corrections. 
The procedure includes the quark-gluon splitting $P_{qg}$ and 
the full implementation of a heavy flavour VFNS\cite{White} and gives 
full LO resummed partons.
Colferai outlined an approach which is also applicable for processes with 
two hard scales. It too 
includes the running coupling but also resummation of 
collinear singularities: 
\begin{equation}
\int \!\!d k^2 (k^2)^{-\gamma-1} K^n(k^2) = \chi^n(\gamma),
\quad \chi^n(\gamma) \,\,\sim \,\, 1/\gamma^{2n+1}, \,\, 
1/(1-\gamma)^{2n+1}\nonumber
\end{equation}
The evolution variable is $s/(QQ_0)$ with conjugate variable $N$. 
Consideration of changes of evolution 
variable to $s/Q^2$ (DIS)
and $s/Q_0^2$ lead to the resummation\cite{Salam} 
$\chi_N^n(\gamma)\sim 1/(\gamma+N/2)^{n+1}, 1/(1-\gamma+N/2)^{n+1}$.
The natural calculations are in DIS scheme with the 
incoming gluon off-shell -- $k^2=Q_0^2 \not=0$. 
Colferai looked at the transformation to the $\msbar$ scheme\cite{Colferai},
showing that the effect is not large.   
Forte presented an approach based on:  
duality -- one has 
$\chi(\gamma(N,\alpha_S),\alpha_S)=N, \gamma(\chi(\alpha_S,M),
\alpha_S)=N$ i.e. the $Q^2$ evolution and $x$ evolution are dual\cite{duality} 
(perhaps not the most important issue); explicit 
imposition of momentum conservation; 
inclusion of the running of the coupling\cite{running};
and symmetrisation, i.e. letting $1/M^n \to 1/(M+N/2)^n$, 
$1/(1-M)^n \to 1/(1-M+N/2)^n$. Overall this leads to a resummed 
$P_{gg}(x,Q^2)$\cite{Forte}. These methods are all rather different in 
details, but now have many similarities, and produce similar results 
for $P_{gg}$ at NLO in the resummation, with a dip below the LO splitting 
function for $x\sim 0.001$, and a slow growth not setting in until 
$x\sim10^{-5}$.  
White also examines phenomenology, at LO
in the resummation only -- 
the impact factor required for $P_{qg}$ is not yet known 
at NLO. The analysis gives a better fit\cite{White} than 
NLO-in-$\alpha_S$ in terms of $dF_2(x,Q^2)/d\ln Q^2$, but the  
enhancement of the 
evolution is too great at small $x$ and the resulting gluon and $F_L(x,Q^2)$
are too small at moderate $x$, suggesting that we need the full NLO 
generalization. Colferai\cite{Colferai} and 
White\cite{White2} both examine the improvement to 
$P_{qg}$. The two approaches are qualitatively similar but  contain 
different higher 
order information: White estimates NLO corrections to the impact 
factor\cite{exact}; Colferai has a resummation of $P_{gg}$ beyond 
NLO via the collinear resummation. Both suggest the effects of the 
NLO resummation beyond fixed order in the quark sector are small
but significant, as seen in Fig.\ref{fig1}.

\begin{figure}[ht]
\vspace{-0.5cm}
\centerline{\hspace{-0.5cm}\epsfxsize=2.5in\epsfbox{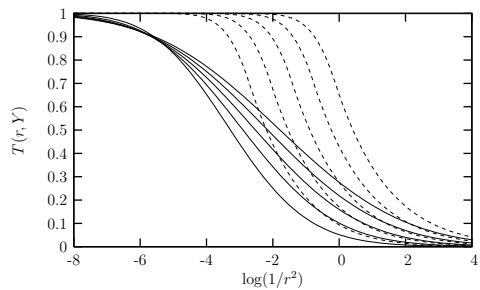}}   
\vspace{-0.2cm}
\caption{Comparison of the dipole cross-section from a fit with only 
saturation corrections (solid) and one which also allows for 
fluctuations (dots).
  \label{fig2}}
\vspace{-0.4cm}
\end{figure}

There was an update on non-linear corrections at very small $x$, 
with various discussions on how to include the saturation corrections, 
and more. 
Soyez considered the nonlinear evolution equation in rapidity 
$Y$ extended to include fluctuations as well as recombination\cite{Soyez1},
i.e. the equation is 
\begin{equation}
\partial_Y \langle T^{(n)} \rangle= \bar\alpha M_{BFKL}\otimes 
\langle T^{(n)}\rangle - 
\bar\alpha M_{BFKL}\otimes\langle T^{(n+1)}\rangle+\bar\alpha
\alpha_S^2 K\otimes\langle T^{(n-1)}\rangle,\nonumber
\end{equation} 
where $T^{m}$ is the amplitude for $m$ dipole scattering.
Hatta demonstrated the origin of the fluctuations via a formal 
derivation of a Bremsstrahlung Hamiltonian\cite{Hatta} 
which gives the 
evolution of $n$-dipole densities. Soyez showed that the 
inclusion of the fluctuations leads to 
a dispersion about the saturation scale\cite{Soyez2}. 
One can be 
sensitive to saturation effects when nominally far away from the saturation 
scale, or conversely, when naively near this scale, fluctuations to even lower
$x$ have no effect since saturation is already reached, while those to higher 
scales reduce saturation effects. The net effect is to    
move the onset of saturation in the dipole 
cross-section considerably downwards, as seen in Fig.\ref{fig2}, which 
compares the dipole cross-section with and without fluctuations. 
Kutak modifies the LO BFKL equation to include a  non-linear 
recombination term\cite{Kutak} 
and also investigates impact parameter dependence\cite{Kutak1}. 
Moreover, high-$x$ effects in the gluon evolution are accounted for 
(though not in the gluon-quark impact factor, i.e. $P_{qg}$).
The impact parameter dependence reduces the effect of 
the nonlinear term and lowers the saturation scale in $Q^2$ for fixed $x$,
i.e. peripheral collisions see smaller densities. 
The conclusion from some rough phenomenology is that we 
``hardly see the effect of saturation'' at 
HERA.

To summarise, there is little agreement in {\it global} fit analyses. 
Not everyone wants to go to NNLO, and not everyone agrees how to do it in 
detail. However, it seems we are now at the stage where NNLO parton 
analyses are essentially complete and reliable and should be done. 
They work a little better than NLO in general. 
There are rather similar results coming from all groups working on small-$x$
resummations to be used in conjunction with fixed order calculations, though
differences in procedures.
The effect of the resummations is moderate until very small
$x$. Empirically, their inclusion can improve the fit a little over NLO. 
There is progress in nonlinear small-$x$ equations, e.g. 
fluctuations, which always seems to be pushing the saturation scale 
lower. It would be good if this approach could match on to higher $x$ 
better  -- at present it is usually confined to some unknown small $x$ region, 
and is missing higher $x$ corrections.
Overall there is lots of improvement in how to calculate using different 
techniques. However, there is 
not enough idea yet where each approach is applicable/needed. We still 
need better (real) phenomenology and, of course, more useful data.

\end{document}